\documentclass[useAMS,usenatbib]{mn2e}
\usepackage{graphicx}
\usepackage{times}

\def\figdir{./Figs}      
\def\etal{et~al.}
\def\d{{\rm d}}
\def\element[#1]#2{{}^{#1}{\rm #2}}
\def\alphaov{\alpha_{\mbox{\scriptsize ov}}}
\def\dov{d_{\mbox{\scriptsize ov}}}
\def\rco{r_{\mbox{\scriptsize co}}}
\def\nablaa{\nabla_{\mbox{\scriptsize ad}}}
\def\nablar{\nabla_{\mbox{\scriptsize rad}}}
\def\nablaov{\nabla_{\mbox{\scriptsize ov}}}
\def\ld{l_{\mbox{\scriptsize d}}}
\def\mco{M_{\mbox{\scriptsize co}}}

\begin{document}
\title[The effect of overshooting in PMS evolution]
      {On the effect of overshooting as predicted by the modelling
       of the pre-main sequence evolution of a 2 solar-mass star}

   \author[J. P. Marques \etal]
   {Jo\~ao P. Marques$^{1,2}$,
    M\'ario J. P. F. G. Monteiro$^{1,3}$,
    Jo\~ao Fernandes$^{2,4}$ \\
    $^1$ Centro de Astrof\'\i sica da Universidade do Porto, 
	 Rua das Estrelas, 4150-762 Porto, Portugal\\
    $^2$ Grupo de Astrof\'\i sica da Universidade de Coimbra,
         Observat\'orio Astron\'omico da Universidade de Coimbra,
         Santa Clara, Coimbra, Portugal\\
    $^3$ Departamento de Matem\'atica Aplicada, Faculdade de Ci\^encias
         da Universidade do Porto, Portugal \\
    $^4$ Departamento de Matem\'atica, Faculdade de Ci\^encias e Tecnologia
         da Universidade de Coimbra, Portugal
	     }

   \date{Accepted 2006; Received 2006; in original form 2005}
   \pagerange{\pageref{firstpage} -- \pageref{lastpage}}
   \pubyear{2006}

\maketitle
\label{firstpage}

\begin{abstract}
  We discuss the effects of convective overshooting in the PMS evolution of intermediate mass stars, by analysing in detail the early evolution towards the main sequence of a 2~$M_\odot$ stellar model.
  These effects can be extremely important in the end of the PMS, when the abundances in CNO elements approach the equilibrium in the centre.
  We provide a possible physical explanation on why a moderate amount of overshooting produces, as the star approaches the ZAMS, an extra loop in the evolutionary tracks on the HR diagram.

  An interesting feature is that there is a very well defined amount of overshooting (for a given stellar mass and chemical composition) beyond which a loop is produced.
  For smaller amounts of overshooting such a loop does not take place and the evolutionary tracks are similar to the ones obtained by \cite{iben65}.
  The amount of overshooting needed to produce the loop decreases with stellar mass.

  We discuss the underlining physical reasons for the behaviour predicted by the evolution models and argue that it provides a crucial observational test for convective overshooting in the core of intermediate mass stars.

\end{abstract}


\section{Introduction}\label{sec:intro}

  Convection and convective overshooting can play a fundamental role in the evolution of a star, by being efficient mixing processes.
  From the coupling of convection with nuclear reactions, significant changes on the evolution may be expected when convective overshoot is present, regardless of using different prescriptions for convection and/or overshooting.
  Several authors have addressed the effect of including overshoot in modelling the evolution of main and post-main sequence stars \cite[e.g.][ and references therein]{stothers90,mowlavi94}.
  But the problem of how much is the amount of overshooting that has to be
included is still far from solved \cite[see][]{renzini87}.
  Several authors modelled core overshooting in stars \cite[e.g.][]{roxburgh78,bressan81,langer86,xiong85,xiong86}, arriving at different results concerning the extent of the convective penetration into the radiative zone, from negligible to quite important.
  Another open issue is how overshoot affects the local temperature stratification \cite[e.g.][]{canuto97}.
  But as far as evolution is concerned the extent of overshooting is the key aspect \citep{deupree00,ribas00}.

  There are several indications, both theoretical and observational, that support the existence of overshooting in the convective core of intermediate and high mass stars.
  The comparisons of mass and radii of eclipsing binaries with theoretical models suggests not only the existence of a certain amount of overshooting, but also its dependence on stellar mass \citep{ribas00}.
   On the other hand theoretical predictions of the apsidal motion rate for very eccentric binary orbits can be compatible with observed values, only if overshooting is included in the models \citep{claret93}.
   Moreover 2D hydrodynamic simulations performed by \cite{deupree00} for stars with masses from 1.2 to 20~$M_\odot$ predict convective core overshooting for all models.
  
  The effect of overshoot on stellar evolution has already been extensively discussed in the literature, but its effect on models of pre-main sequence stars has never been discussed taking the full network of nuclear reactions into account. 
  The overall effect to be expected is a slight delay in the evolution, redefining the age at which the star arrives at the zero age main sequence (ZAMS).
  As the zero point for age of young stars is in itself an open problem, such a global effect of overshoot on the PMS has been ignored.

  The mixing at most of the overshoot layer at the border of a convective core in main sequence stars is expected to be efficient \cite[e.g.][]{browning04}.
  This should also be the case for the end of the PMS evolution, as the time scales for mixing due to convection are still much smaller than the time scales for nuclear reactions.
  The uncertainties on the thermodynamic stratification within an overshoot layer are not expected to be of major relevance for determining the effects on the early evolution, but the extent of the overshoot layer and its effect on mixing the stellar material at that location are bound to be decisive.

  In a previous work \citep{marques04} we studied the effects of 
the time step and convective overshooting in the calibration of the young binary EK~Cep.
  It has been shown that the time step adopted for the PMS evolution is a key numerical aspect to adequately reproduce the expected behaviour in this rapid phase of evolution.
  When the time step is properly defined for calculating the evolution of intermediate mass stars with overshoot the evolutionary tracks near the end of the PMS display an extra loop.
  This loop seems to be present in the output of previous works \cite[e.g.][]{siess00} although it has never been discussed in the literature, as for most cases an inadequate time step in the integration procedure would mask the loop as being a numerical artifact.
  Consequently its nature and physical validity has never been analysed or discussed in the literature.

  Given that an adequate definition of the time step for the PMS evolution together with the most recent set of physics produces the loop near the ZAMS (when there is a significant amount of overshooting), it has lead us to perform a detailed analysis on the underlying reasons why such a loop is predicted by an evolutionary code.
  In particular the goal of this work is to establish if such an effect should be expected in real stars.
  Instead of dismissing it as a numerical artifact, as implicitly done by other authors, we have investigated if it could be consistent with the physical picture that the new state-of-the-art physics predicts (in particular the network of nuclear reactions).
  Here we provide such an analysis showing that there may be a physical reason for such a behaviour.
  In spite of the difficulties of modelling convection and convective overshoot, our analysis seems to indicate that the loop in the HR diagram at the end of the PMS evolution could happen in real stars. 

  We start by discussing how overshoot is introduced in a one dimensional evolutionary code and move on to show evolutionary tracks without and with overshoot for a particular case of a 2~$M_\odot$ star.
  In order to illustrate the physics behind the existence of the loop at the end of the PMS we then discuss the limit case of a specific value of the extent of the overshoot and the evolution of the chemical abundances that produce the loop.
  The paper ends by addressing how different input physics and stellar parameters change the limit value of the overshoot producing the loop and what observational evidence could be expected in order to identify a star undergoing such a phase in its PMS evolution.

\section{Modelling convective overshooting in PMS evolution}
\label{sec:overs}

  In this work we propose to study in detail the effects of convective overshooting in the PMS evolution of intermediate mass stars, as they approach the ZAMS, and so it is relevant to start by addressing how overshooting is implemented in an evolutionary code.
  We will restrict our discussion to the case of intermediate mass stars (around 2~$M_\odot$), where there is a small central convective core.

\begin{figure}
\centering
\begin{tabular}{cc}
\includegraphics[width=\hsize]{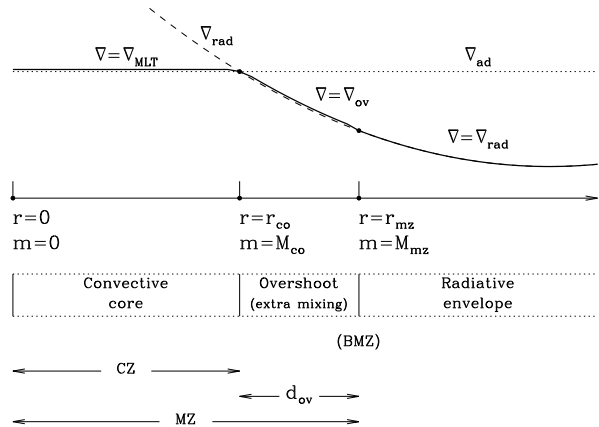}
\end{tabular}
\caption{Schematic representation of the behaviour of temperature gradient (thick continuous line) in the convective core and overshoot region of an intermediate mass star.
  The notation is defined in the text.
  CZ represents the proper convective zone, while MZ is the mixing zone.}
  \label{fig:nablas_ov}
\end{figure}

  The boundary of a convective core $\rco$, with a mass $\mco$ (see Fig.~\ref{fig:nablas_ov}), is defined by the instability criteria, namely the Schwarzschild criterium or the Ledoux criterium if the chemical composition is not homogeneous.
  For the Schwarzschild criterium the border between the convective zone (core) and the radiative zone (envelope) is located where the adiabatic gradient $\nablaa$ equals the radiative gradient $\nablar$.
  Convective elements that reach the boundary with non-zero velocities overshoot it, reaching a region where $\nablar{<}\nablaa$.
  To transport all the energy generated in the convective core upwards, the luminosity transported by radiation in the zone immediately above the boundary must be superior to the total luminosity, that is, in this zone one must have $\nablar{<}\nablaov{<}\nablaa$.
  
  All models of overshooting have inadequacies \citep{renzini87,canuto97}, as it is not possible to fully describe in one dimension the effect of overshooting plumes.
  The standard option used is to parametrize the effect overshoot may have in the local structure by adopting a prescription.
  There are two major aspects of overshoot that we need to consider in such a prescription if we want to discuss the effect on stellar evolution.
  These are the temperature stratification within the overshoot layer and the mixing taking place in such a region at the top of the ``proper'' convective core.

\subsection{Thermal stratification at the overshoot layer}
\label{subsec:ovt}

  The temperature stratification below $\rco$ is given by the MLT formulation (which corresponds to the convective zone; CZ).
  Given the very small superadiabaticity expected in the proper convection zone the actual structure is not sensitive to the details of the particular formulation we use, or to the mixing length parameter $\alpha$.
  Here we will adopt a value of $\alpha{=}1.35$.
  As shown by \citet{marques04} for stars with masses around 2~M$_\odot$ any other reasonable value (as the one obtained from a solar calibration) will produce the same evolutionary tracks near the end of the PMS evolution.

  As discussed in several works, the temperature stratification in the overshoot layer is expected to be subadiabatic.
  We assume, as in \citet{y205}, that the stratification in the overshoot layer is unaffected by the overshoot, a result also obtained by \citet{browning04}.
  Consequently, within the overshoot layer of thickness $\dov$
(corresponding to $\rco{\le}r{\le}\rco{+} \dov$) the stratification is assumed to be radiative ($\nablaov{=}\nablar$).
  Although it is a crude choice, its implementation in an evolutionary code is simple and does not affect the key aspects of overshoot that can be expected to interfere with the nuclear reactions that take place as the star approaches the main sequence.
  To confirm if this is so we also consider here other options for the thermal stratification of the overshoot layer.

  The representation of overshooting should be implemented using more sophisticated non-local formulations or the results of 3D simulations
\cite[e.g.][]{grossman96,canuto97,browning04}.
  Unfortunately, most of these formulations can not be fully implemented in detailed evolutionary codes as the non-local character of overshooting compromises the efficiency of the integration in time for stellar evolution.

\subsection{Mixing in the overshoot layer}
\label{subsec:ovc}

 The key assumption, that we adopt here, is that the mixing of the stellar matter within the overshoot layer is highly efficient taking place at the same time scales as in the proper convection zone \cite[see][]{browning04}.
  That implies that in terms of evolution, and in particular at the end of the PMS, it is fully justifiable to consider that convection and overshooting induce instantaneous mixing as far as evolution is concerned.
  Also, if the stellar mass is not too high, as is the case for the regime of intermediate mass stars we are addressing here, such mixing takes place in time scales much shorter than the mean lifetimes of all reactions in the CNO bi-cycle that may take place before the star reaches the ZAMS.

  As long as the assumption of instantaneous mixing within the overshoot layer stands, the implications for the evolution of more complex formulations for the thermal stratification of overshoot are equivalent to the results provided by the simpler formulation that is being adopted.
  For the phase of evolution we discuss here the classic assumption of using the limit of instantaneous mixing is adequate as it provides a representation of a lower limit for the extent of overshoot that may produce the loop.
  This extent is well below the expected extent of overshoot for this mass range \citep{ribas00}.

  In order to establish to what extent our results depended on the mixing model in the overshoot region, we also consider models calculated using the prescription of \cite{ventura98}.
  Accordingly, the evolution of element $i$ follows the diffusion equation:
\begin{equation}
\frac{d X_i}{dt} =
  \left(\frac{\partial X_i}{\partial t}\right)_{\mbox{\scriptsize nuc}}
  + \frac{\partial}{\partial m_r}\left[\left(4 \pi r^2 \rho\right)^2 D \;
  \frac{\partial X_i}{\partial m_r}\right] \;.
\end{equation}

  The diffusion coefficient $D$ is approximated in a convective zone by $D{=}u\,\ld/3$, where $u$ is the average turbulent velocity, computed according to eqs (88), (89) and (90) of \cite{cgm96}, and $\ld$ is the convective scale length.
  The convective scale length $\ld$ is given by 
\begin{equation}
\ld = \frac{z_{\mbox{\scriptsize up}} z_{\mbox{\scriptsize low}}}
   {z_{\mbox{\scriptsize up}} + z_{\mbox{\scriptsize low}}} \;,
\end{equation}
where $z_{\mbox{\scriptsize up}}$ is the distance from the top of the convective zone increased by $\beta H_P^{\mbox{\scriptsize top}}$ and analogously for $z_{\mbox{\scriptsize low}}$.
  Here, $\beta$ is a fine tuning parameter, necessary for an exact fit of actual stars, and $H_P^{\mbox{\scriptsize top}}$ the pressure scale height at the top of the convective region.
  The parameter $\beta$ is constrained by $\beta{<}0.25$.

  For diffusive overshooting, we write:
\begin{equation}
u = u_{\mbox{\scriptsize b}} \;
  \exp \pm\left[\frac{1}{\zeta f_{\mbox{\scriptsize thick}}} \;
   \ln \left(\frac{P}{P_{\mbox{\scriptsize b}}}\right) \right],
\end{equation}
where $u_{\mbox{\scriptsize b}}$ and $P_{\mbox{\scriptsize b}}$ are the turbulent velocity and the pressure at the border of the convective zone, $P$ is the local pressure, $\zeta$ is a free parameter and $f_{\mbox{\scriptsize thick}}$ is the thickness of the convective region in fractions of the local $H_P$.
  This way, the turbulent velocity decreases exponentially outside convective regions.
  The parameter $\zeta$ controls the e-folding distance with which the velocity of convective eddies decay outside convective regions; a higher $\zeta$ means that the velocity of the convective eddies decays slower and therefore a bigger region is affected by partial convective mixing.
  The diffusive scale is approximated by $\ld{=}\beta H_P$ in overshoot regions.

\subsection{Extent of the overshoot layer}
\label{subsec:d_ov}

  Following the assumptions discussed above the only aspect remaining to be defined is the extent $\dov$ of the overshoot layer.
  We adopt the common prescription (see \citealt{maeder75}; \citealt{mowlavi94}; and \citealt{y205}) that parametrises the extent of the overshooting layer in units of the local pressure scale height $H_P{\equiv}|\d r/\d\ln P|$.
  So, the extension of the overshooting is given by
$  \dov {=} \alphaov {\cdot} {\mbox{Min}}\left(H_{P},\rco\right)$,
  where the border $\rco$ of the ``proper'' convective core
corresponds to the radius defined by the Schwarzschild criterium
and $\alphaov$ is a free parameter to be defined.
  We note, however, that eclipsing binaries data indicates that this parameter is dependent on stellar mass \citep{ribas00}, reaching a value of about 0.2 for $M{\simeq}2~M_\odot$.
  As discussed above, inside $r_{\mbox{\scriptsize mz}}{\equiv}\rco{+}\dov$ there is complete mixing (we shall call this zone the mixed zone - MZ).

\section{Overshooting and evolution of the internal structure}
\label{sec:evol}

\begin{table}
\caption{Mean lifetimes (in years) for the nuclei involved in the
CNO bi-cycle, for $X{=}0.7$ and $\rho{=}64$~g~cm$^{{-}3}$
(from \citealt{caughlan62}).
  Mean lifetimes for secondary nuclei are not included since their $\beta$-decay is very fast ($10^2$ -- $10^3$~s) compared to the mean lifetimes of the primary nuclei ($T_6{\equiv}T/10^6$~K).
} \label{tab:tau}
  \begin{center}
\begin{tabular}{lcc}
\hline
\noalign{\smallskip}
  \qquad\quad Reaction & $T_6=20$  & $T_6=15$ \cr
\noalign{\smallskip}
\hline
\noalign{\smallskip}
$\element[12]{C}~ 
  [\element[12]{C}(\element[1]{H},\gamma)\element[13]{N}]$ &  
  $1.47\times 10^4$ & $2.96\times 10^{6\phantom{1}}$ \\
$\element[13]{C}~
  [\element[13]{C}(\element[1]{H},\gamma)\element[14]{N}]$ &  
  $3.62\times 10^3$ & $7.43\times 10^{5\phantom{1}}$ \\
$\element[14]{N}~
  [\element[14]{N}(\element[1]{H},\gamma)\element[15]{O}]$ &  
  $1.73\times 10^6$ & $1.82\times 10^{8\phantom{1}}$ \\
$\element[15]{N}~
  [\element[15]{N}(\element[1]{H},\element[4]{He})\element[12]{C}]$ &
  $7.06\times10^1$ & $2.64 \times 10^{4\phantom{1}}$ \\
$\element[15]{N}~
  [\element[15]{N}(\element[1]{H},\gamma)\element[16]{O}]$ &
  $1.69 \times 10^5$ & $6.19 \times 10^{7\phantom{1}}$ \\
$\element[16]{O}~ 
  [\element[16]{O}(\element[1]{H},\gamma)\element[17]{F}]$ &  
  $1.09\times 10^8$ & $6.48\times 10^{10}$ \\
$\element[17]{O}~
  [\element[17]{O}(\element[1]{H},\element[4]{He})\element[14]{N}]$ &  
  $3.38\times 10^6$ & $6.94\times 10^{10}$ \\
\noalign{\smallskip}\hline
\end{tabular}
\end{center}
\end{table}

\begin{figure*}
\centering
\includegraphics[width=14cm,angle=0]{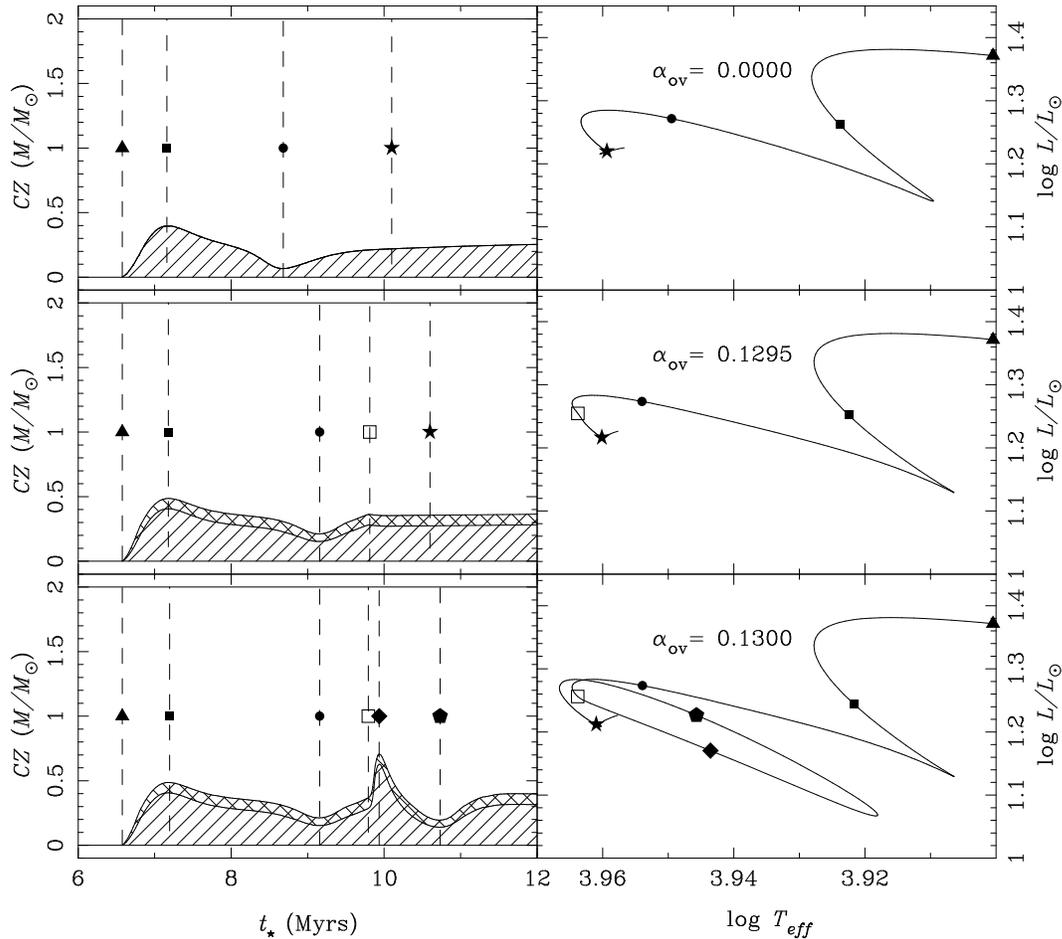}
\caption{Left panels: evolution of the convective zones for models with different values of $\alphaov$: $\alphaov{=}0$ (top panels), $\alphaov{=}0.1295$ (middle panels) and $\alphaov{=}0.1300$ (bottom panels).
  Cross-hatched areas represent the extent of overshooting.
  Right panels: evolutionary tracks in the HR diagram for the same cases as in the left panels.
  Symbols on the track correspond to evolutionary ages of special interest:
they are also shown in the left panels (see Table~\ref{tab:times}).
} \label{fig:hrcz}
\end{figure*}

  Models of a 2~$M_{\odot}$ star with values of $\alphaov$ varying from 0 to 0.13 have been produced (see Table~\ref{tab:times}). 
  The evolution was calculated using an initial helium abundance $Y_{0}{=}0.28$, initial metallicity $Z_{0}{=}0.02$ and mixing length parameter $\alpha{=}1.35$.
  Models were calculated using the CESAM stellar evolutionary code \citep{morel97}.
  The up-to-date physical ingredients and adopted numerical procedures relevant for the early evolution near the ZAMS - used to build the models discussed here - are described in detail by \citet{marques04}.

  The key aspect in the calculation of models near the ZAMS is the energy production.
  For the mass range discussed here this is the CNO bi-cycle which consists of two cycles.
  In the first, the CN cycle, only carbon and nitrogen are involved:
  \begin{eqnarray}
\element[12]{C}+\element[1]{H} &\rightarrow& \element[13]{N}+\gamma
\nonumber\\
\element[13]{N} &\rightarrow& \element[13]{C} + e^{+}+\nu_e \nonumber\\
\element[13]{C}+\element[1]{H} &\rightarrow& \element[14]{N}+\gamma
\nonumber\\
\element[14]{N}+\element[1]{H} &\rightarrow& \element[15]{O}+\gamma
\nonumber\\
\element[15]{O} &\rightarrow& \element[15]{N} + e^{+}+\nu_e \nonumber\\
\element[15]{N}+\element[1]{H} &\rightarrow& 
\element[12]{C}+\element[4]{He}.\label{rq:cno}
  \end{eqnarray}
  The second, the ON cycle, is entered by the termination of the last reaction of (\ref{rq:cno}) through the $\gamma$ channel (instead of the $\alpha$ channel):
  \begin{eqnarray}
\element[15]{N}+\element[1]{H} &\rightarrow& \element[16]{O}+\gamma
\nonumber \\
\element[16]{O}+\element[1]{H} &\rightarrow& \element[17]{F}+\gamma
\nonumber \\
\element[17]{F} &\rightarrow& \element[17]{O} + e^{+}+\nu_e
\nonumber \\
\element[17]{O}+\element[1]{H} &\rightarrow&
\element[14]{N}+\element[4]{He}.\label{rq:cno1}
  \end{eqnarray}
  Table~\ref{tab:tau} shows, however, that the termination of the fusion of $\element[15]{N}$ through the $\alpha$ channel is much more likely (about $10^{4}$ times) than through the $\gamma$ channel.
  The equilibrium timescale of the CN cycle is therefore much shorter than that of the full CNO cycle; for the PMS phase, only the processes in (\ref{rq:cno}) - the CN cycle - will be considered in our discussion (although the models use the full network).
  There is still another ON cycle, which is entered by the termination of the last reaction of (\ref{rq:cno1}) (the fusion of $\element[17]{O}$) through the $\gamma$ channel; its importance, however, is very small.
  Table~\ref{tab:tau} also shows that the slowest process in (\ref{rq:cno}) is, by far, the combustion of $\element[14]{N}$.
  Therefore, before the abundances in CN elements reach equilibrium almost all original $\element[12]{C}$ and $\element[13]{C}$ must be burned into $\element[14]{N}$.

\begin{table}
\caption{Age (in Myrs) of the models at different phases of evolution for different values of $\alphaov$.
These phases correspond to: a -- the central MZ appears; b -- the central MZ reaches the first maximum; c -- minimum extent of the central MZ; d -- ZAMS.
Also indicated is the type of symbol used in Fig.~\ref{fig:hrcz} to locate these phases.
} \label{tab:times}
\begin{center}
\begin{tabular}{ccccc}
\hline
  Phase & \multispan3 $\alphaov$ & Symbol \\[4pt]
  & 0.0000 & 0.1295 & 0.1300 & (Fig.~\ref{fig:hrcz}) \\
\hline
a  & \phantom{1}6.6 & \phantom{1}6.6 & \phantom{1}6.6 & {\it black triangle} \\
b  & \phantom{1}7.2 & \phantom{1}7.2 & \phantom{1}7.2 & {\it black square} \\
c  & \phantom{1}8.7 & \phantom{1}9.2 & \phantom{1}9.2 & {\it black circle} \\
   & - & \phantom{1}9.7 & \phantom{1}9.7 & {\it open square} \\
   & - & - & \phantom{1}9.8 & {\it black diamond} \\
   & - & - & 10.5 & {\it black pentagon} \\
d  & 10.1 & 10.6 & 12.1 & {\it star} \\
\hline
\end{tabular}
\end{center}
\end{table}

  All models discussed here use the full network of reactions (PP chains and CNO cycle) to calculate the energy production.
  The PMS birthline \citep{palla91} is not used in these calculations as the chemical profile at the core of the models for a 2~$M_\odot$ star near the ZAMS are not significantly changed by what happens at the very early stages of stellar evolution.

  Figure~\ref{fig:hrcz} shows evolutionary tracks for several values of $\alphaov$.
  The most remarkable feature is that there are two kinds of tracks, with or without a ``loop'' just before the ZAMS.
  The transition between these two types of tracks happens at a very definite value of $\alphaov$: for $\alphaov{=}0.1295$ there is no ``loop'' in the evolutionary tracks on the HR diagram, while there is a ``loop'' for $\alphaov{=}0.1300$.
  For values of $\alphaov$ lower than 0.1295, the tracks do not differ significantly; the same happens for values of $\alphaov$ higher than 0.1300. 
  This loop can also be seen in the evolutionary models available online, by \citet{siess00}, for a 2~$M_{\odot}$ star with $Z{=}0.02$ and $\alphaov{=}0.2$.
  This feature requires an explanation.
  In the following sections we will describe in detail the evolution of the internal structure of a 2~$M_{\odot}$ star without and with overshooting in order to understand the origin of these features in the models of PMS evolution.

\subsection{Evolution without overshooting}
\label{subsec:evol_nov}

  The evolutionary track of a 2~$M_{\odot}$ star on the HR diagram without overshooting is shown in Fig.~\ref{fig:hrcz} (upper panel), as well as the evolution of the convective zones (top right panel), during the last phase of the PMS.

\begin{figure}
\centering
\includegraphics[width=\hsize]{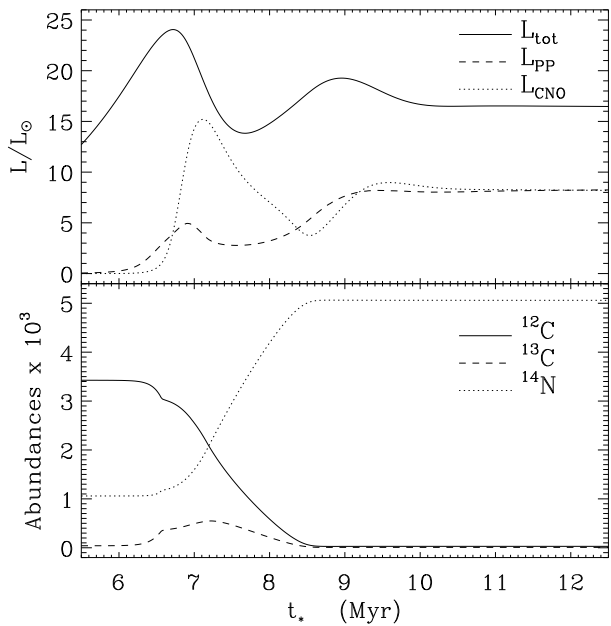}
\caption{Evolution of the production of energy (top panel) and
the central abundances (lower panel) without overshooting.
} \label{fig:abund_no-ov}
\end{figure}

  At an age $t_{\star}{=}5$~Myrs, the star is completely radiative and all its layers are contracting.
  The central temperature rises, until it becomes high enough to burn $\element[3]{He}$ and $\element[12]{C}$.
  The combustion of both $\element[3]{He}$ and $\element[12]{C}$ depends strongly on the temperature ($\varepsilon{\sim}T^{16}$ to $T^{18}$); therefore, the energy generated is strongly concentrated in the centre of the star.
  The energy flux from the central regions becomes so high that it can not be transported by radiation alone.
  A central convective zone appears at this point.
  Figure~\ref{fig:abund_no-ov} shows the evolution of the production of energy and  the central abundances of CN elements.
  Comparing Figs.~\ref{fig:hrcz} (top left panel) and \ref{fig:abund_no-ov}, it is clear that the convective core appears at the time when $\element[12]{C}$-burning starts (at an age $t_{\star}{=}6.6$~Myrs); this instant is indicated in Fig.~\ref{fig:hrcz} by the black triangle (see Table~\ref{tab:times}).

  The structure of the core implies that $\rho_{\mbox{\scriptsize c}}/\overline{\rho}$ is about 6 for a convective model, while being higher than 50 for a radiative one.
  Because the convective model has to be less centrally concentrated, the central zone must expand when convection appears; this expansion absorbs practically all the energy produced in the centre by $\element[12]{C}$-burning, which does not, thus, reach the surface.
  We have the apparently paradoxical result that when nuclear energy sources become important the total luminosity of the star decreases (see top right panel of Fig.~\ref{fig:hrcz}), although there is now a new energy source.

  As the burning of $\element[12]{C}$ becomes more efficient with rising central temperature and density, the central convective zone grows; when the abundance of $\element[12]{C}$ in this zone decreases, the energy generated in the centre of the star decreases as well. 
  In Fig.~\ref{fig:hrcz}, the black square indicates the instant the central convective zone reaches its maximum extent ($t_{\star}{=}7.2$~Myrs).   
  As the energy generated in the centre of the star decreases, the extent of the central convective zone decreases.
  As discussed previously, when this happens the region of the convective zone must now strongly contract in order to adapt to the structure of a radiative core.
  The central density rises again, as well as the central temperature.

  For a 1~$M_{\odot}$ star, PP chains would generate now practically all the energy, since $\element[14]{N}$-burning would be always too slow for the CN 
cycle~(\ref{rq:cno}) to generate a large fraction of the energy (see 
Table~\ref{tab:tau}, for $T_6{=}15$).
  For a 2~$M_{\odot}$ star, however, PP chains can not stop the contraction of the stellar core.
  As almost all $\element[12]{C}$ originally in the core has been burned and the convective core retracts as a consequence, the central zone contracts again. 
  The luminosity generated by this contraction causes the luminosity of the star to increase, as can be seen in the top right panel of Fig.~\ref{fig:hrcz} and in Fig.~\ref{fig:abund_no-ov}.
  As the central temperature rises, the combustion of $\element[14]{N}$ becomes more efficient (Table~\ref{tab:tau}); again, the energy generation becomes concentrated in the centre of the star; the convective zone grows and, as a consequence, the luminosity of the star decreases again.
  The instant when the convective zone reaches a minimum is indicated in Fig.~\ref{fig:hrcz} by the black circle ($t_{\star}{=}8.7$~Myrs).
  The contribution of the CNO cycle~(\ref{rq:cno}) to the energy generation grows, until the star is finally in hydrostatic equilibrium. 
  The ZAMS is indicated in Fig.~\ref{fig:hrcz}, top right panel, by the star ($t_{\star}{=}10.1$~Myrs). 

  Figure~\ref{fig:abund_no-ov} shows that the abundances of $\element[12]{C}$ and $\element[13]{C}$ do not grow significantly while the $\element[14]{N}$-burning reaction is becoming more efficient; all $\element[12]{C}$ produced by the burning of $\element[14]{N}$ very rapidly burns into $\element[13]{C}$, which even more rapidly burns into $\element[14]{N}$ again (Table~\ref{tab:tau}).
  The burning of $\element[14]{N}$, because it is so slow compared with the other reactions of the CN cycle~(\ref{rq:cno}), acts as a bottleneck. 

\subsection{Evolution with overshooting}
\label{subsec:evol_ov}

  As seen above, for $\alphaov{\ge}0.1300$ there is a loop in the evolutionary track in the HR diagram during the final stages of the PMS.
  For $\alphaov{\le}0.1295$ there is no such loop; the transition is quite abrupt. 
  Figure~\ref{fig:hrcz} shows the evolution of the convective zones with $\alphaov{=}0.1295$ (middle panels); it is qualitatively similar to the evolution with $\alphaov{=}0$.
  The two main differences are that with $\alphaov{=}0.1295$ the extent of the MZ is higher and the evolutionary time is longer (see Table~\ref{tab:times}).
  The extent of the MZ at the minimum after the main $\element[12]{C}$-burning is also significantly higher.
  These differences are easily explained by the extra extent of the MZ caused by the overshooting; the burning of $\element[12]{C}$ takes more time because the MZ extends further and therefore there is more $\element[12]{C}$ to burn, while the rate of burning is similar with $\alphaov{=}0.0$ and $\alphaov{=}0.1295$ (since the central densities and temperatures are similar).
  Table~\ref{tab:times} shows the ages of the models at some stages of the evolution.
  Before the appearance of the central MZ, the evolutionary times are the same.

  Figure~\ref{fig:hrcz} shows the evolution in the HR diagram (bottom right panel) and the evolution of the convective zones (bottom left panel) of a model with $\alphaov{=}0.1300$.
  In the bottom right panel of Fig.~\ref{fig:hrcz}, the black triangle shows the instant the central MZ appears for the first time ($t_{\star}{=}6.6$~Myrs); this zone reaches its maximum extent around $t_{\star}{=}7.2$~Myrs (black square).
  As the $\element[12]{C}$ present in the central MZ burns out, the MZ retreats, reaching a minimum at $t_{\star}{=}9.2$~Myrs (black circle).
  It grows again as the burning of $\element[14]{N}$ becomes more efficient, and, while with $\alphaov{=}0.1295$ the reactions of the CNO cycle (\ref{rq:cno}) reach equilibrium and the central MZ stabilises, with $\alphaov{=}0.1300$ the MZ grows suddenly. 
  Until this point, the evolution for $\alphaov{=}0.1295$ and $\alphaov{=}0.1300$ is the same; here, they split completely.
  This event (at $t_{\star}{=}9.7$~Myrs) is indicated in Fig.~\ref{fig:hrcz} with the open square.
  The central MZ grows rapidly, reaching a maximum (${\simeq}0.7~M_{\odot}{=}0.35~M_{\star}$!) at $t_{\star}{=}9.8$~Myrs (black diamond).
  This sudden growth of the MZ is caused by an increase in the central abundance of $\element[12]{C}$ and $\element[13]{C}$, which are burned very efficiently because the central temperature is now higher than during the main $\element[12]{C}$-burning phase; so, the energy produced in the centre of the star increases rapidly.

\begin{figure}
\centering
\includegraphics[width=\hsize]{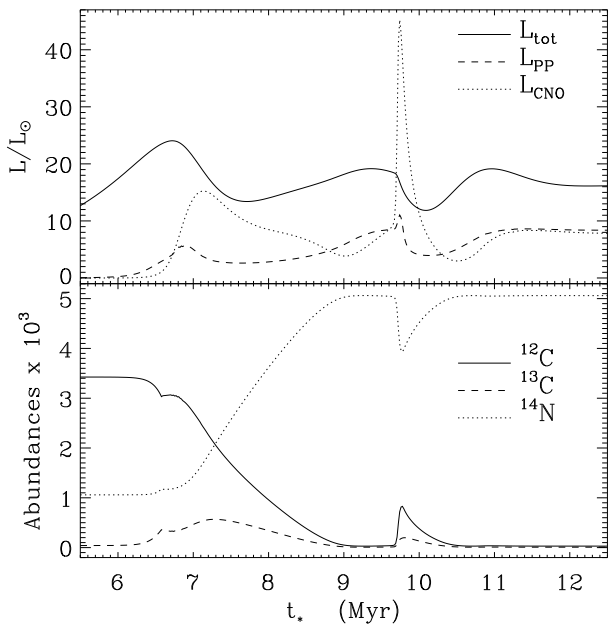}
\caption{Evolution of the production of energy (top panel) and central abundances (lower panel) with $\alphaov{=}0.1300$.
}
  \label{fig:abund_ov-1300}
\end{figure}

  Figure~\ref{fig:abund_ov-1300} shows the evolution of the production of energy and the central abundances with $\alphaov{=}0.1300$; the main difference to Fig.~\ref{fig:abund_no-ov} ($\alphaov{=}0$) is the sudden rise in the abundance in $\element[12]{C}$ and $\element[13]{C}$ (as well as a drop in the abundance in $\element[14]{N}$).
  This causes the rise in the production of energy through the CNO cycle seen in the left panel of Fig.~\ref{fig:abund_ov-1300}, causing in turn the growth of the central MZ seen in Fig.~\ref{fig:hrcz}.
  From this point, the MZ retracts again, reaching a new minimum at $t_{\star}{=}10.5$~Myrs (black pentagon).
  The model of the star then returns to the ``normal'' track, reaching the ZAMS at $t_{\star}{=}12.1$~Myrs.

  The MZ retracts so fast because it grew so fast before.
  This sudden growth causes an expansion of the central regions, which leads to a big drop in the central temperature and particularly a big drop in the central density.
  The reaction rates of $\element[12]{C}$ and $\element[13]{C}$-burning, depending so strongly on the temperature and density, drop rapidly.
  Since this sequence of events (after the sudden growth of the central abundance in $\element[12]{C}$ and $\element[13]{C}$) does not depend on the extent of the central MZ when $\alphaov{\ge}0.1300$, depending only on the central temperatures and densities, the format of the loop on the HR diagram does not depend on the actual extent of overshooting above $\alphaov{=}0.1300$.

\subsection{The central abundances}
\label{subsec:expl}

  In this section we study the evolution of the central abundances for $\alphaov{=}0$ and $\alphaov{=}0.1295$.
  The value $\alphaov{=}0.1295$ is particularly interesting because it is almost enough to produce a ``loop'' in the evolutionary tracks on the HR diagram.
  The comparison between the cases $\alphaov{=}0$ and $\alphaov{=}0.1295$ shows the reasons why a ``loop'' is produced for $\alphaov{\ge}0.1300$ and why is this transition so abrupt.

\begin{figure*}
\centering
\includegraphics[width=\hsize]{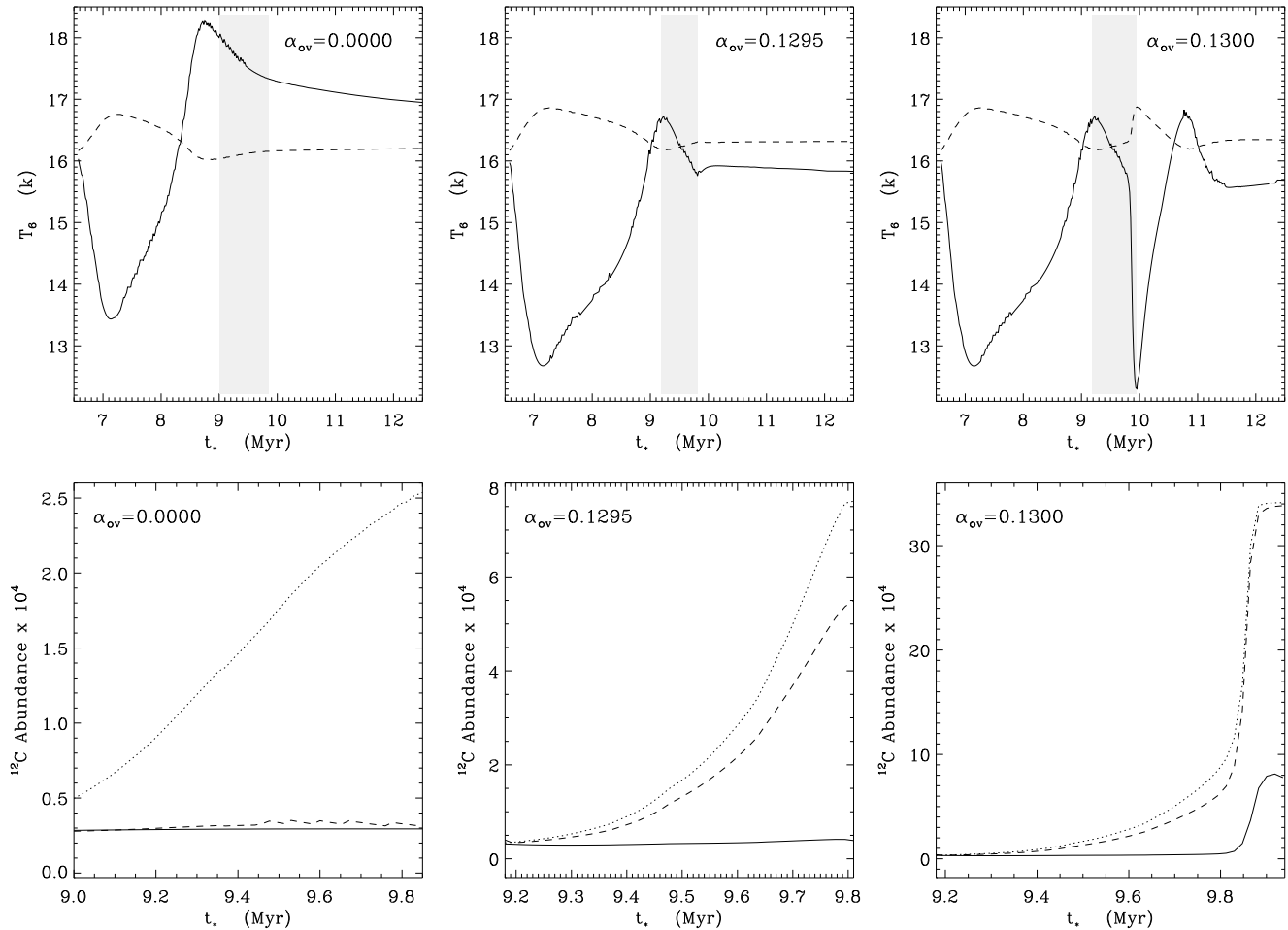}
\caption{Upper panels: evolution of the temperature of the BMZ (full line) for several values of $\alphaov$ (where $T_6{\equiv}T/10^6$).
  The temperature ($T_{\rm ign}$) at which the mean lifetime for $\element[12]{C}$ is $0.75$~Myrs (for the density at the BMZ), is also shown as a dashed line.
  Lower panels: the abundance of $\element[12]{C}$ the BMZ finds during the final expansion of the MZ (dashed line); evolution of the central abundance in  $\element[12]{C}$ during the same period (full line); and the abundance in $\element[12]{C}$ the BMZ would find if there would be no depletion of $\element[12]{C}$ above the MZ due to nuclear reactions (dotted line).
  The shadow in the upper panels indicates the time range of the corresponding lower panel.
}
  \label{fig:tc_c12}
\end{figure*}

  The chemical composition of $\element[12]{C}$ at the central convective zone reaches a minimum at $t_{\star}{=}8.7$~Myrs for $\alphaov{=}0$ and at $t_{\star}{=}9.1$~Myrs for $\alphaov{=}0.1295$.
  The $\element[12]{C}$ is not fully burned above the MZ at this stage.
  If temperatures above the MZ remained too low to burn significant amounts of $\element[12]{C}$ there during the final expansion of the MZ, the MZ would add this $\element[12]{C}$ during the following expansion.
  Figure~\ref{fig:tc_c12} (upper panels) shows the evolution of the temperature of the BMZ (labelled $T_{\mbox{\scriptsize BMZ}}$) for $\alphaov{=}0,0.1295,0.13$.
  The temperature ($T_{\mbox{\scriptsize ign}}$) at which the mean lifetime for $\element[12]{C}$ is $0.75$~Myrs (for the density at the BMZ), is also shown.
  It is clear that the temperature at the BMZ decreases as the amount of overshooting increases; overshooting extends the MZ, so the BMZ is farther from the centre of the star, where the temperature is lower.
  Without overshooting, the temperature of the BMZ is higher (during the final expansion of the MZ) than the temperature needed to significantly deplete $\element[12]{C}$ before the MZ adds it.
  So we expect that the MZ adds almost no $\element[12]{C}$ during its final expansion.

  As we increase the amount of overshooting, the temperature at the BMZ decreases (see upper panels of Fig.~\ref{fig:tc_c12}), making $\element[12]{C}$-burning above the MZ incomplete.
  As a result, some $\element[12]{C}$ is added to the MZ during its expansion.
  Finally, when $\alphaov{=}0.1295$ there is almost no depletion of $\element[12]{C}$ above the MZ during its last expansion.
  The increase with $\alphaov$ of the amount of $\element[12]{C}$ added to the MZ when $\alphaov$ approaches the critical value of 0.1295 is very high.
  Even a small increase of $\alphaov$ from $\alphaov{=}0.1290$ to $\alphaov{=}0.1295$ is enough to increase noticeably the amount of $\element[12]{C}$ added to the central MZ.

  If there is no $\element[12]{C}$ added to the core (as it is the case without overshooting), the CNO reactions reach equilibrium shortly after $\element[14]{N}$-burning becomes efficient enough (that is, shortly after the central MZ starts to grow for the last time).
  When, due to the extra extent of the central MZ caused by overshooting there is more $\element[12]{C}$ added to the MZ than that produced by $\element[15]{N}$-burning, there must be an excess of $\element[12]{C}$ (and $\element[13]{C}$)-burning to keep the relative abundances of CN elements at equilibrium.

\begin{figure}
\centering
\includegraphics[width=\hsize]{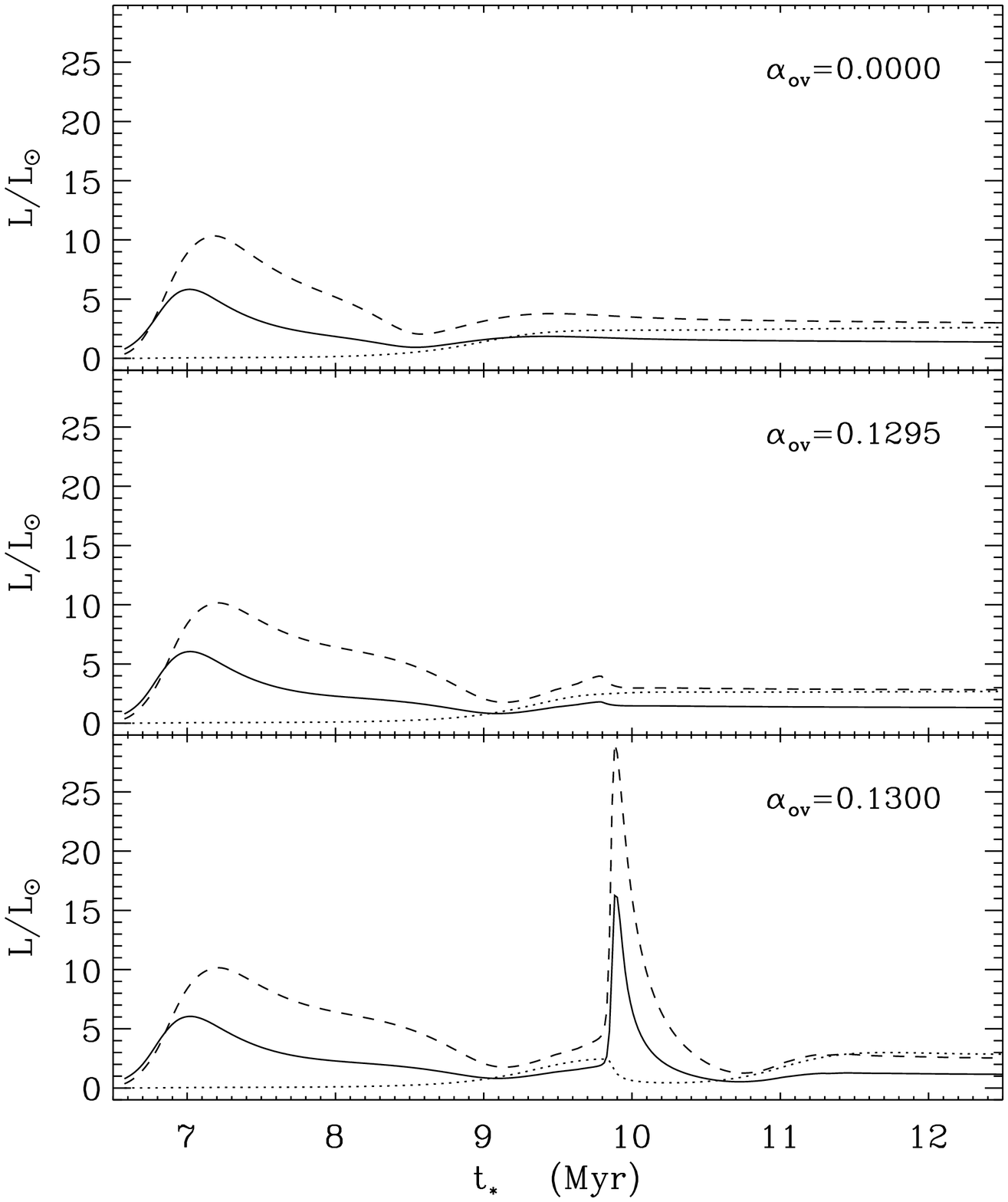}
\caption{Evolution of the luminosity produced within the central MZ by the burning of  $\element[12]{C}$ (full line),  $\element[13]{C}$ (dashed line) and  $\element[14]{N}$ (dotted line). 
  The panels correspond to different values of $\alphaov$.
} \label{fig:ls}
\end{figure}

  This excess burning of $\element[12]{C}$ (and $\element[13]{C}$, since the extra burning of $\element[12]{C}$ generates an excess in the abundance of $\element[13]{C}$ relative to the equilibrium abundances) causes an increase on the luminosity produced within the central MZ; to transport this extra luminosity out, the central MZ must grow.
  Figure~\ref{fig:ls} shows that only when $\alphaov$ approaches the critical value does this increase in the luminosity becomes noticeable.
  The growth of the central MZ must bring in turn an even greater amount of $\element[12]{C}$ to the MZ, causing an even greater increase in the luminosity produced within the central MZ.
  This is only stopped because such a fast increase in the energy produced within the core causes it to expand and cool, making CNO reactions much less efficient.

  In short, the abruptness of the transition from an evolution with a loop to an evolution without a loop comes from the sensibility of the CNO reactions to the temperature, as the parameter $\alphaov$ determines the temperature of the BMZ, which in turn determines how much $\element[12]{C}$ is burned before the MZ grows.

\section{Sensitivity of the results to the input physics}

  The physics of the early evolution for a 2~$M_\odot$ star is expected to be simpler than for the more advanced stages of evolution.
  Aspects as diffusion and settling as well as the existence of steep chemical gradients as a result of hydrogen burning in the main sequence are not important in the cases discussed here.
  In this section we address some of the key aspects that may be relevant and the possible effect these may have on the existence of the loop in real stars.

  Although we presented a very definite value of $\alphaov$ beyond which a loop is produced in the evolutionary tracks on the HR diagram, this ``critical'' value can not be taken at face value, since it depends on several known and unknown factors.
  Such a critical value depends mainly on the nuclear reaction rates, but other aspects of the models may modify the actual number.
  We stress that the main goal of this work is not the determination of a critical value for $\alphaov$ but to establish what are the physical and numerical ingredients responsible for producing the loop in the evolution calculation.

\subsection{Physics and numerics of the overshoot layer}
\label{subsec:phy-ov}

  Numerical simulations and observational tests (see Section~\ref{sec:intro}) consistently indicate that stars are expected to have mixing zones that go beyond the border predicted by an instability criterium for convection.
  The unresolved questions are the extent and the physics of these extra mixing zones.
  Here we have included such an extra mixing zone (overshoot layer) at the top of a small convective core by adding an extra mixing layer of size $\alphaov H_p$ to the proper convection zone modelled according to the mixing length theory.

  There are different prescriptions for overshooting as discussed in Section~\ref{sec:overs} which will provide slightly different results for the same value of $\alphaov$.
  In order to confirm that our results are not affected by the thermal stratification adopted within the overshoot layer we have also calculated the evolution when the limit of $\nablaov{=}\nablaa$ is used within the overshoot region.
  As long as instantaneous mixing is assumed the critical value of $\alphaov$ does not change by more than about 0.0005.
  The implications for the evolutionary track in the HR diagram are not changed and the existence of the loop is still confirmed in this limit.
  The actual stratification in the overshoot layer of stars is expected to lie between these two values ($\nablar{<}\nablaov{<}\nablaa$).
  Consequently, any temperature profile will produce the same behaviour we discuss in this work.

  Models calculated with different precision (number of shells for each model and time step of the evolution - see \citealt{marques04}) or using different numerical methods yield different values for the critical $\alphaov$.
  These differences (below 0.001) are a direct consequence of the need for a careful numerical scheme to follow in time and space the border of convective and mixing zones in stellar interiors.
  These numerical tests indicate that the existence of the loop is to be expected when adequate numeric procedures are used for time and space integration.
  As far as this work is concerned these uncertainties are small corrections which will only become relevant if the measured extent of overshoot in stars is close to the critical value.
  As discussed above (Section~\ref{sec:intro}), observational constraints on the value of the extent of an overshoot layer are not clear yet.
  Some works \citep{ribas00} indicate that the amount of overshooting in the cores of 2~$M_{\odot}$ stars can be significantly larger than $\alphaov{\sim}0.13$.

\subsection{Diffusive versus instantaneous mixing}
\label{subsec:phy-mix}

  Here we have adopted instantaneous mixing for the overshoot layer.
  This is a crucial assumption that needs to be confirmed.

  The time scale of the relevant nuclear reactions is significantly larger than the expected convective turnover time for the core.
  However, the mixing at the overshoot layer, in particular at the very top, may have longer mixing timescales when compared with the proper convective zone.
  Models incorporating overshoot as a diffusion process \citep{ventura98} indicate that for stars with convective cores the effect on the mixing is equivalent to an instantaneous overshooting around 0.2~$H_p$.
  This result has been reinforced by \cite{ventura05} where nuclear burning and mixing were self-consistently coupled for post-main sequence stars and compared with instantaneous mixing.
  The confirmation that such a behaviour is also valid for the pre-main sequence evolution is required, but given the physics at play in this early phase no major differences are expected.

  In order to verify this assumption we have also considered the case when mixing is not instantaneous outside convective zones.
  However, the exponential decay of the velocities outside the convective zone means that over time the mixed zone is larger.
  During the first retreat of the central CZ extra $\element[12]{C}$ is brought to the core so the energy generation there is higher than it would if no overshooting were present.
  That keeps the convective zone bigger and the temperature at its border lower, so when the CZ grows again it finds above a significant abundance of $\element[12]{C}$, as happens in the case of instantaneous mixing in the overshoot zone.

  Consequently, a loop is also produced on the evolutionary tracks calculated using the prescription of \cite{ventura98} with values of $\zeta {\ge} 0.01$ and $\beta {\ge} 0.05$.
  These values should be compared to $\zeta{=}0.03$ used by \cite{ventura05} to fit the cluster NGC 1866.
  The loop is found to be very similar to the one obtained with instantaneous mixing; both the trajectory in the HR diagram and the time it takes.

\subsection{Stellar rotation}
\label{subsec:phy-rot}

  Rotation can also play a role on how overshoot behaves.
  Considering the aim of this work, the important questions to be addressed are the following: can rotation change, considerably, the time scale of mixing when compared to the burning time scale?
  Is the efficiency of mixing affected by 
the rotation?

  \cite{browning04} presented three-dimensional simulations of core convection for a 2~$M_\odot$ star.
  They found that the core has a differential rotation and an overturning period of the global scale convection of about one month.
  On the other hand they support that overshooting and penetrative convection are both effective in mixing the chemical composition. 
  The existence of an efficient overshooting layer for rotating stellar cores was also found by \cite{deupree98} using 2D simulations for a 8.75~$M_\odot$ star.
  Consequently there is no evidence to support (on the contrary) that rotation will affect the results reported here, and in particularly our assumption concerning the instantaneous mixing.

  It is worth noticing that \cite{meynet00} found for high mass stars ($9{\le}M/M_\odot{\le}120$) that rotation modifies the tracks in the HR diagram as moderate overshoot would do.
  A similar conclusion is reached by \cite{noels04} for A stars, where the authors argue that rotation reduces the efficiency of convection and consequently the extent of overshooting.
  However the observational evidence \citep{herbst05} indicates that most PMS stars loose their angular momentum well before arriving at the main sequence, when they are already slow rotators.
  In such cases suppression of overshooting due to rotation is weaker and so a loop may be expected in intermediate mass PMS slow rotators close to the main sequence.

\subsection{Dependence on stellar parameters}
\label{subsec:phy-par}

  The critical value of the parameter $\alphaov$ depends strongly on the stellar mass (or the size of the convective core), decreasing for higher stellar masses.
  For a 3~$M_{\odot}$ star, there is a loop in the evolutionary track on the HR diagram for $\alphaov{\ge}0.0593$, while for a 4~$M_{\odot}$ star the same thing happens for $\alphaov{\ge}0.0391$.
  So, for higher mass stars a relatively small amount of overshooting is enough to produce a loop before the ZAMS.
  This is due to the faster evolution of higher mass stars, leaving less time for the depletion of $\element[12]{C}$ in the regions above the MZ before the MZ grows again.
  So $T_{\mbox{\scriptsize ign}}$, which we defined as the temperature at which the mean lifetime of $\element[12]{C}$ is $\tau_{\mbox{\scriptsize C12}}{=}0.75$~Myrs, should be replaced by the temperature at which $\tau_{\mbox{\scriptsize C12}}{=}0.27$~Myrs for a 3~$M_{\odot}$ star and $\tau_{\mbox{\scriptsize C12}}{=}0.13$~Myrs for a 4~$M_{\odot}$ star.
  The region where the temperature is equal to $T_{\mbox{\scriptsize ign}}$ lies therefore nearer the centre of the star, and a relatively small amount of overshooting is enough to bring the BMZ to those regions.

  The evolution is faster for higher masses; the loop lasts about $4{\times}10^5$~yrs for a 3~$M_{\odot}$ star and about $2{\times}10^5$~yrs for a 4~$M_{\odot}$ star, making a possible observation much more difficult. 
  An additional difficulty for the case of a 4~$M_{\odot}$ star is that the birthline (see \citealt{palla91,palla92}) is closer to the ZAMS; the loop should happen at an age of about $5{\times}10^5$~yrs after the birth of the star.
  The effects of the protostar phase could be of some importance at such an early age.
  Also, the assumption of instantaneous mixing becomes weaker as we move towards higher mass models.
  In such cases the requirement to consistently follow the mixing and the nuclear reactions becomes more important, being necessary to calculate models with time dependent mixing coupled with nuclear burning (as done in \citealt{ventura05}, for post-main sequence stars) at the central core of these young stars.

\section{Expected observational evidences}

\begin{figure}
\centering
\includegraphics[height=\hsize,angle=-90]{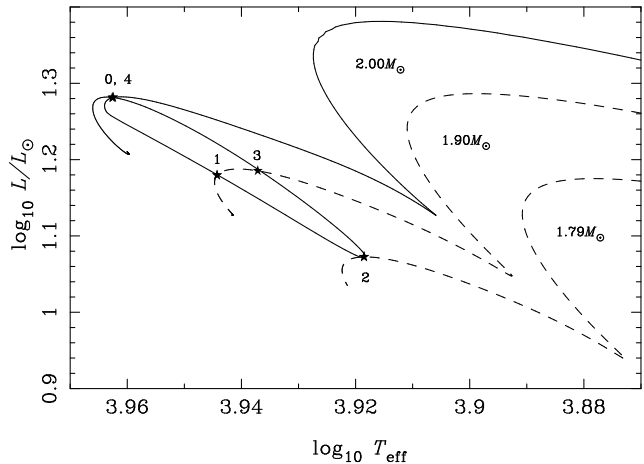}
\caption{Evolutionary tracks in the HR diagram for three stars.
  One with $M{=}2~M_{\odot}$ (full line) having a loop due to the presence of overshoot, and two stars of smaller mass (dashed lines) without overshoot which cross the track of the more massive star.
  The point ``0'' marks the beginning of the loop, the point ``4'' (in the same location on the HR diagram) marks the end.
} \label{fig:loop_hr}
\end{figure}

  The observational confirmation that a particular star is in the loop can be straight forward, but the probability of finding such a star is very small as it depends on the relative time that a star spends in the loop.
  Very young clusters with stars in the range of 1.5 to 3.0~$M_{\odot}$ arriving at the ZAMS are preferential targets for finding loop stars.
  The location at the HR diagram (see Fig.~\ref{fig:loop_hr}) has to be complemented with the determination of the mass of the star.
  Consequently the study of possible candidates in binary systems, with a smaller mass component to determine the age precisely, is the natural observational test for the existence and determination of the characteristics of loop-stars.

  An alternative to study binaries, is to obtain seismic data on several individual stars in order to discriminate between stars with the same radius but different masses.
  Such a test can be applied as long as a few oscillations can be identified in order to provide the observational value for the large frequency separation $\Delta_\ell$ for modes of degree $\ell$ and calculated at a reference frequency $\nu_{\mbox{\scriptsize r}}$ (e.g. \citealt{monteiro02,fernandes03}).
  This quantity (for $\ell{=}0$) can differ by as much as 6~$\mu$Hz between a loop-star and a star without overshoot at the same location in the HR diagram.

  If very precise seismic data are available the small frequency separation, $\delta_{\ell,\ell{+}2}$, can also be determined using modes of degree $\ell$ and $\ell{+}2$ (being calculated at the same reference frequency).
  The small frequency separation, being sensitive to the central regions of the star (e.g. \citealt{monteiro02b}), can differ by as much as 30\% at $\nu_{\mbox{\scriptsize r}}{=}1200~\mu$Hz for location 2 in Fig.~\ref{fig:loop_hr}, providing a powerful additional test to the large frequency separation.

  Another seismic indicator, mainly sensitive to the central regions of the stars \citep{roxburgh03,roxburgh05}, is the ratio $r_{02}{\equiv}\delta_{02}/\Delta_1$ of the small separation (using modes of degree 0 and 2) and the large separation (for modes of degree 1).
  This quantity can provide an even more definite discriminant for stars with strong structural differences in the interior but with the same radius.
  This is the case between loop-stars and the correspondent star without overshoot at the same location in the HR diagram.
  The values of $r_{02}$ for models occupying the same location in the HR diagram show differences that can be as high as 0.025 ($\sim$26\%).

\section{Conclusions}

  We have shown that a moderate amount of overshooting in evolution models of young intermediate mass stars can cause a loop in the evolutionary tracks on the HR diagram just before the ZAMS.
  This is a consequence of using a more precise numerical scheme for calculating the time-step together with the best up-to-date physics.
  The loop corresponds, for a 2~$M_{\odot}$ star, to a drop of about 35\% in luminosity and about 1000~K in effective temperature; the loop lasts about 1.5~Myrs.
  A detailed description on why the full nuclear network predicts in the models such a loop has been presented.
  This analysis indicates that such a behaviour may be expected if the standard formulations for core overshoot in stellar models are representative of the average role convective overshooting has in the early evolution of intermediate mass stars.
  The major effect of the existence of the loop is on the age of the star at the ZAMS, being higher by as much as 20\% relative to the evolution without the loop.

  The observational identification of loop-stars can provide definite tests on some of the key aspects of the physics that determine the evolution in early and main sequence intermediate mass stars.
  In particular the existence and extent of overshoot at the core of these stars and the reaction rates for the branches in the CNO cycle responsible for producing the loop.
  Such a combination of precise observational constraints make the existence of loop-stars an interesting and worth pursuing observational opportunity to test the modelling of stellar structure and evolution at these early stages.
  The observational challenge is the short duration of the loop; less than about 2~Myrs.
  The loop is even shorter for higher stellar masses, for which the amount of overshooting needed to produce a loop becomes smaller and therefore more likely.
  Further analysis on how overshooting and nuclear reactions are coupled in the higher mass models is required in order to identify if the loop can be expected in this regime.

  With the forthcoming asteroseismic space missions \cite[e.g.][]{baglin03} and ongoing ground based campaigns for the seismic study of stars across the HR diagram, there will be a large set of young stars whose seismic properties will be determined with sufficiently high precision to be possible to use the seismic analysis we have discussed here.
  There are in particular campaigns planned for the seismic observation of young clusters whose ages are estimated to be around the expected values for the existence of loop stars (from about 5 to 12~Myrs).
  Consequently, the possibility of confirming the existence of loop stars may be possible in the not so distant future.
  The implications of finding a few such stars in a young cluster will be of great relevance for the modelling of the early evolution of intermediate mass stars.

\section*{Acknowledgements}

  The authors are grateful to an anonymous referee for valuable comments and suggestions that help to improve the manuscript.
  This work was supported in part by the Portuguese {\it Funda\c{c}\~ao para a Ci\^encia e a Tecnologia} through grant (JPM) {\scriptsize SFRH/BD/9228/2002}, and projects {\scriptsize POCI/CFE-AST/55691/2004} (MJM) and {\scriptsize POCI/CTE-AST/57610/2004} (JPM) from POCI, with funds from the European programme FEDER. 
  This work has been performed using the CESAM stellar evolution code available at  {\small\tt http://www.obs-nice.fr/cesam/}


\label{lastpage}
\end{document}